\documentclass{appolb}
\usepackage{graphicx}
\def\KL{K\"all\'en-Lehmann\,\,}
\begin{document}
\title{Physics after the discovery of the Higgs boson%
\thanks{Presented at the Final HiggsTools meeting, Durham UK, September 2017.}%
}
\author{J. J. van der Bij
\address{Institut f\"ur Physik\\
Albert-Ludwigs-Universit\"at Freiburg}
}
\maketitle
\begin{abstract}
I  show that with the discovery of the Higgs boson we have entered a new phase of 
our understanding of nature. This leads us towards a paradigm shift in the search for
possible new physics, away from major extensions like supersymmetry and towards minimalistic
extensions that largely preserve the structure of the standard model. To discover such new physics,
precision may be more important than energy. Precision = Discovery! A possible path for the future
is sketched.
\end{abstract}
  
\section{Introduction}
It is clear to everyone that with the discovery of the Higgs boson a milestone has been reached.
However from the theoretical point of view, finding the Higgs boson is not much of a discovery.
After all, without the Higgs boson, the theory would not be quantum mechanically consistent. Of course no one
seriously disputes the validity of quantum mechanics. Quantum mechanics is, philosophically speaking, simply
a better theory than classical mechanics, because it contains a fundamental unit of angular momentum, thereby
eliminating one dimensionful parameter from the theory. For the same reason, relativity is better than classical
mechanics, because it eliminates velocity as a dimensional parameter. One therefore gets closer to the ideal
that a theory of nature should be a theory of pure numbers. Pure QCD would be such a theory.
Knowing the mass of the Higgs boson however, is important as is the fact that nothing else has been found.
Therefore we are naturally lead to a paradigm-shift, as I mentioned in my congratulory letter to the CERN Director General.\\

\begin{quote}
Dear Fabiola,\\
I want to congratulate CERN for the great running of the machine and 
the brilliant work of the detectors. I think the results are great, I have 
rarely seen such a convincing null-experiment. I am sure this will lead to 
the long overdue paradigm-shift away from the view ``the standard model is 
wrong and we will have to see what is beyond" and towards the view
``we know the standard model is true and we have to understand why".\\
In the attachment I give you my answer [1-3].\\
good luck, Jochum
\end{quote}

\section{Why the standard model}
There are three steps one can take in natural philosophy.\\

The first step is the observation, that there actually are rules and regularities,
which one can call natural laws. It took humanity a while to reach this step in a conscious way.
Without this one is in a realm of pre-rational thought, where events are determined by the will
of gods, demons or fairies. Within popular non-scientific culture this realm is still very present.
With the realization that regularities are present, numbers and precision already entered the field.
One should mention here Nabu-rimanni from Babylon, whose astronomical tables were not improved until
 the discovery of the telescope [4].

The second step is to determine what the laws of nature are. Here the invention of the scientific method
was crucial, consisting of systematic measurements, guided by mathematical calculations and vice versa.
With the discovery of the Higgs boson and the absence of signs of new physics at the LHC, we should now
take the view that nature has spoken and we know that the standard model is true.

This brings us to the third step. We want to understand why the laws are the way they are and not different.
As Einstein put it: I am interested to know whether God had a choice when He created the world.
This is to be interpreted in the following way: are there a few fundamental principles that
allow one to determine the laws of nature in a unique way. This sort of question is often considered to be beyond
the range of present day physics. However tentative partial progress has been made recently.
One assumes the following principles [1,2]:
\begin{enumerate}
\item Gravity is a geometrical theory.
\item The quantum fields describing matter should be consistent with any
    compactification of gravity.
\end{enumerate}
Using gravitational anomalies [5] the following results can then be derived:\\
\begin{enumerate}
\item The fundamental gauge group of nature is SU(5). 
\item There must be exactly three generations including right-handed neutrinos. 
\item After symmetry breaking the standard model is the only possible low energy theory.
\end{enumerate}
These results apply only to the chiral sector of the theory.
For a popularized derivation there is a podcast [3].

\section{Historical interlude}
I will give here a personal view of the development in physics since I entered the field.
In 1978 the Utrecht-Paris-Rome triangle meeting took place in Utrecht.
At this meeting it was already clear that the standard model was basically right.
However people were not happy about this, because it contains too many parameters.
Therefore principles, none of them very rigorous, were invented as a guidance to go beyond the standard model.
At the time people were still optimistic and had the hope to maybe calculate the quark masses.
It is interesting to see what is left of these ideas now, after almost 40 years of experiment.
I remember three proposals.

 First there was the idea of naturalness by G. 't Hooft [6].
This says that parameters can be small, when one increases the symmetry by putting the parameters to zero.
This was attractive as, at the time, the masses of the known fermions were small and they could be 
considered as small corrections to some underlying larger dynamics.\footnote{Now we know that the top mass is heavy, the idea becomes less attractive.}
In this sense,  all parameters in the standard model are natural with the exception of the Higgs boson mass,
which should be infinite because of quadratic divergences.  In subsequent developments, the quadratic divergence
of the Higgs mass has come to be considered as the major problem of the standard model. This is somewhat strange,
as for instance the presence of three generations appears to be a more obvious problem. Anyway the argument was used to
construct ever more fanciful models like technicolor, extra dimensions and supersymmetry, often in combination.
These models in the end contained even more parameters than the standard model and had to be fine-tuned to extreme precision
not to get into trouble with the data. Even renormalizability was thrown out of the window.
All this is gone now. However what is left are the 't Hooft anomaly matching
conditions, that roughly say that anomalies in an effective low-energy theory should correspond to the anomalies
of the fundamental theory. This can be understood on the basis of dispersion relations [7,8].

The second idea came from M. Veltman, who was guided by simplicity, however allowing possibly for strong
interactions. The idea was that the muon could be a very strongly bound state between an electron and a scalar singlet.
This does not work in detail, as there are strong limits on fermion structure. However singlets can surely be present in the theory [9]
and tend to help for instance in solving cosmological problems. Somewhat later he suggested to cancel
 quadratic divergences within the standard model [10].

The third idea came from J. Iliopoulos, who suggested that things should be determined by renormalization group equations
and an infrared fixed point would play a role [11]. In a peculiar twist of fate the renormalization group running of the couplings
does seem to play a role, but in an opposite way. Running the equations up to the Planck scale one finds that one is close to
a flat potential precisely near the Planck scale. This helps with inflation, but what it actually means is anyone's guess.
However it is the only direct connection there is between the Planck scale and the weak scale within the standard model.

So altogether these ideas are still tentative and have not brought us much further. The real and secure progress has been
in ever more difficult and precise calculations, but these will not make you famous and are not everyone's cup of tea.

\section{Minimalistic extensions}
\subsection{Definition}
We can summarize the results after the measurements at the LHC as follows.
There are no particles beyond the standard model observed at the LHC, 
nor is there  new flavour physics seen at the LHC.
Furthermore the Higgs boson has been found and there is a general agreement between 
precision data and standard model predictions. This has strong implications for the possibilities of new physics.

The absence of flavour effects at the LHC puts limits on the scale of new physics of $O(10\,\rm{TeV})$.
This implies that major changes to the fundamental structure of the standard model can only exist at scales beyond
the range of the LHC. No reasonable theoretical models of this type exist. 
Also the general agreement of the electroweak precision data  with the standard model implies that major changes 
are not possible. 

Therefore extensions must be minimalistic so they do not effect the fundamental
structure of the standard model. This leaves few possibilities.
Examples are
inert scalar multiplets, that do not couple to fermions,
or non-chiral fermions. These are both good candidates for dark matter.
One can also consider so-called St\"uckelberg Z'-bosons that only couple to the standard model through mixing
with the hypercharge vector-boson.

But the simplest ``safe" extensions are of course singlets.\\
It is reasonable to expect singlet fields to be present in the
scalar sector, after all they exist in the fermion and in the 
gauge sector.
Moreover they are the extensions of the standard model with the
smallest number of parameters.
Since singlets do not change the basic gauge structure of the standard
model, it is a matter of taste whether such extensions still belong
to the standard model. One could call this the non-minimal standard model (NMSM). \\

Is there anything one can say here about the possible new physics for which there is evidence?
There is clear evidence for the existence of dark matter. Dark matter poses an interesting question.
Is it just like ordinary matter, but simply does not couple to photons? This so-called WIMP scenario
can easily be described by minimalistic models. The alternative is that dark matter is something completely
unrelated to the standard model; a very light Bose-Einstein condensate is a recently popular candidate.
A second possibility for which there is some evidence is a sterile neutrino. As such neutrinos are singlets
they obviously belong to the minimalistic extensions. The third evidence for new physics is $(g-2)_{\mu}$.
Here I simply cannot find a reasonable explanation; the measured effect is very large. Any new physics one introduces 
to explain this, tends to get into trouble elsewhere. The new $g-2$ experiment at Fermilab should clarify the situation.

\subsection{Lepton non-universality}
The importance of the results at the LHC is that the Higgs boson mass has been determined
and that no new particles, carrying weak charges appear to exist. This implies that we can 
compare theory predictions with data to a much higher level of precision than before. Precise predictions
in the theory are sensitive to radiative corrections, dependent on the Higgs boson mass. 
Before the discovery of the Higgs boson, the data were used to constrain the range of the mass for the Higgs boson.
Now that all parameters of the model are known the theory predictions are essentially exact, so one can 
look for much smaller deviations than before. To look for possible deviations in the precision data,
we consider a model with $n$ neutral sterile fermions, that only mix with the neutrinos of the standard model.
Such particles can play a role in cosmology, i.e. in leptogenesis or as dark matter candidates.
The consequence is that the Pontecorvo-Maki-Nakagawa-Sakata ($PMNS$) matrix is part of a more general mixing matrix.

Taking into account the standard model neutrinos and the extra neutrinos we find that
the mass eigenstates $( \nu_1 \cdots  \nu_{3+n} )$
and flavour basis ($\alpha= e,\mu,\tau$): $\{\nu_i=\nu_{L_\alpha},\,N_n\}$ are connected by a unitary $(3+n)\times(3+n)$ matrix:\\

\begin{displaymath}
\left(\begin{array}{c} \nu_1 \\ \vdots \\ \nu_{3+n} \end{array}\right) =  \left(\begin{array}{cc} PMNS & {\cal W} \\ {\cal W}^\dagger & {\cal V} \end{array}\right) \left(\begin{array}{c} \nu_{L_e} \\ \vdots \\ N_n\end{array}\right)\,.
\end{displaymath}
As a consequence the
$PMNS$ matrix, being a  submatrix, is not necessarily unitary.
We describe the deficit from unitarity by the $\epsilon$ parameters:

\begin{displaymath}
\epsilon_\alpha = \sum_{i>3} |{\cal U}_{\alpha i}|^2 = 1- \sum_{\beta= e, \mu, \tau }|{\cal U}_{\alpha \beta}|^2\,.
\end{displaymath}

As a consequence low energy parameters are affected by the $\epsilon$ parameters.
For example the Fermi constant in muon-decay is modified by the following relation:\\ 

\begin{displaymath}
G_\mu^2 = G_F^2 (1-\epsilon_e)(1-\epsilon_\mu)\,,
\label{eq-Gmu}
\end{displaymath}
with $G_\mu$ the Fermi constant  measured in muon decay, and $G_F$  the Fermi parameter, derived from the standard model theory
without $\epsilon$ parameters.

Other corrections appear in meson-decays and in precision measurements at LEP; we are therefore in the
lucky position that we can combine low-energy and high-energy (LEP) measurements. 
When we do this, we find that the most precise data cannot be well fitted to the model [12,13], even allowing
for the presence of the $\epsilon$ parameters. The origin of the problem was tracked to a single measurement,
namely the forward-backward asymmetry of bottom quarks at LEP. This measurement would lead to a  large and unphysical negative value
for $\epsilon_e$. The other measurements are in good agreement with each other, leading to a value
$\epsilon_e \approx 2.10^{-3}$, excluding $\epsilon_e =0$ at the 2-3~$\sigma$ level. 
A number of experiments that can test this result are underway, new measurements of $\sin^2_{\theta, eff}$, an improvement
on $M_W$, meson decays in $b$ and $\tau$ factories, the ratio $W\rightarrow e$/$W\rightarrow \mu$ and a precise
lattice evaluation of $f_{\pi}$. In combination these could lead to a 5$\sigma$ discovery.

\subsection{Spectral densities}
Let us first discuss why one needs a Higgs field in the first place. It is normally said that one needs a Higgs field 
in order to generate a mass for the vector-bosons in a gauge-invariant way. This is strictly speaking not true.
The theory of massive vector-bosons alone is simply the unitary gauge of a gauged non-linear sigma model, so 
gauge invariance is not the problem here. The problem is that the non-linear sigma model is a theory with constraints.
As in classical Lagrange mechanics such constraints correspond to infinite forces, that lead to problems in the quantum mechanics
of the theory. The effect can be mimicked by studying the standard model in the limit of an infinite Higgs boson mass. This limit is 
interesting as on the one hand the Higgs particle is infinitely heavy and on the other hand it is strongly interacting.
We are therefore faced with the classical problem: what happens when an irresistible force hits an immovable object.
The answer: logarithmic quantum corrections.\footnote{That is at one loop; at higher loops powers of $m_H^2$ appear.}
 This explains why precision measurements at low energy could constrain the mass of the Higgs
boson. This is in contrast to atomic physics which does not constrain the mass of the muon.

This said, it means that the real reason we need the Higgs field is renormalizability.
This however does not imply, that one must have a single Higgs particle peak.
Fundamental quantum field theory tells us only that the Higgs field must
have a \KL spectral density [14,15]. This density can be largely arbitrary, but must fall off
fast enough at infinity, since otherwise the theory is not renormalizable. 
Since in some sense the Higgs field is considered to be different from other fields,
it is not unreasonable to expect a non-trivial density. The premier scientific goal
regarding electroweak symmetry breaking is thus to measure the \KL spectral density
of the Higgs propagator. In praxis this means measuring the Higgs line-shape (width)
and looking for further peaks with a smaller than standard model signal strength.

This is all a bit abstract and one might wonder whether there is a reason to expect more than a single Higgs particle peak.
Indeed, the value of the mass of the Higgs boson in combination with the mass of the top quark is intriguing.
Running the couplings up to the Planck mass one finds that the Higgs potential becomes unstable, but just barely so.
A slightly heavier Higgs $m_H=129.4\,\rm{GeV}$ would lead to a flat potential at the Planck scale [16]. This would help in
explaining inflation. If the spectrum does not consist of a single particle peak
one can increase the average value of $m_H^2$ and get a stable Higgs potential.
The simplest model that can do this is the Hill model [17], which has the following Lagrangian in
the Higgs sector.

\begin{displaymath}
{\cal L}  = -\frac{1}{2} (D_\mu \Phi)^{\dagger}(D_\mu \Phi) - \lambda_1/8
(\Phi^\dagger \Phi -f_1^2)^2
  -\frac{1}{2}(\partial_\mu H)^2 - \frac{\lambda_2}{8} (2 f_2 H - \Phi^\dagger \Phi)^2,
\end{displaymath}
where $\Phi$ is the Higgs field and $H$ is  the Hill field, a real scalar singlet.
Notice that the theory has no ${H^4}$ coupling; it is essentially a pure mixing model.
This is consistent with renormalizability.
So the model describes two Higgs bosons behaving like a standard model Higgs boson but with reduced couplings.
The Higgs propagator becomes,
$$D_{HH}(k^2) = \frac {\sin^2 \alpha}{k^2 + m_+^2} + \frac {\cos^2 \alpha}{k^2 + m_-^2}.$$
This is sufficient to study Higgs signals experimentally (interaction basis).
It corresponds to a \KL spectral density of two $\delta$-peaks.
If one takes $\cos^2(\alpha)\,m_-^2 + \sin^2(\alpha)\,m_+^2 \geq (129.4\,\rm{GeV})^2$, one can stabilize the Higgs potential
at the Planck mass.

The generalization to more fields is straightforward.
One takes 
$n$ Hill-fields, $H_i$, with couplings, $g_i$, that satisfy the 
sum rule, $$\Sigma g_i^2 = g^2_{Standard~model}.$$
This can be generalized to a continuum.
$$\int  \rho (s) ds  = 1,$$
where $\rho$ is the \KL density from the textbooks.\\

An interesting \KL spectral density can be generated by assuming that the Hill field $H$
moves in more than four dimensions [18-20], which can be taken to be infinite and flat.
We call such models HEIDI models, because of the german pronunciation of high-D(imensional).
In this case one is led to the following propagator,

\begin{eqnarray*}
\label{higgsprop4}
D_{HH}(q^2)=\Bigg(q^2 +M^2-\frac{\mu^{8-d}}{(q^2+m^2)^{\frac{6-d}{2}} \pm \nu^{6-d}} \Bigg)^{-1}.
\end{eqnarray*}
In this expression $d$ is the number of dimensions which should satisfy $d\leq 6$, in order to insure
renormalizability. Actually $d$ does not necessarily have to be integer to have a proper propagator.
The parameter $\mu$ describes the mixing of the higher-dimensional Hill field with
the standard model Higgs; indeed putting $\mu=0$ one gets the ordinary Higgs propagator with Higgs boson mass $M$.
The parameter $m$ is a higher-dimensional mass term and the parameter $\nu$ describes the mixing between higher
dimensional modes. Depending on the parameters this propagator describes zero, one or two peaks plus a continuum.
The continuum would correspond to a part of the Higgs field moving away in the extra dimensions; experimentally this would
be interpreted as an invisible decay. The HEIDI models can also stabilize the Higgs-potential [21].
For example one could have a 90\% standard model Higgs at $125\,\rm{GeV}$, a 5\% standard model like Higgs at $142\,\rm{GeV}$ and a 
5\% invisible continuum with an average Higgs mass-squared around $(180\,\rm{GeV})^2$. This would lead to a flat Higgs potential at the Planck mass.
Therefore it is important to measure the properties of the $125\,\rm{GeV}$ Higgs boson as precisely as possible, in particular the overall
cross-section normalized to the standard model is of interest.  One should also look for further standard model-like Higgs bosons.
In these models, the branching ratios to standard model particles are 
the same as in the standard model for a Higgs boson with the same mass. The invisible continuum might be hard to see.
Conclusion:\\
 Higher dimensions may be hidden in the Higgs line-shape!
\vskip 0.4cm
\begin{center}
{\it \large Where~is~Heidi~hiding~?}\\
\vskip 0.2cm
{\it \large Heidi~is~hidden}\\
\vskip 0.2cm
{\it \large in~the~high-D~Higgs~Hill~!}\\
\end{center}

\section{Beyond the LHC: a Higgs factory}
The discussion in the previous section has made clear that just finding the Higgs boson is not
sufficient to completely establish the standard model, as important information can be hidden in the Higgs boson line-shape.
Therefore one needs a Higgs factory to address this question. This will have to be a lepton machine.
The situation is somewhat similar to the situation with the $Z$-boson. The $Z$-boson was discovered
at a hadron collider. However to determine its width, more precisely the line-shape, a lepton collider (LEP) was needed.
This measurement led to the important information, that there are three light neutrinos.
Similarly measuring the line-shape of the Higgs boson is needed to give information on the possible presence of
hidden higher dimensions. The question is then what kind of a lepton collider is needed and how well can one do.

One could consider a muon collider, but at the moment this seems to be science fiction.
Present discussions focus only on the direct production of a standard model Higgs-boson at $125\,\rm{GeV}$.
However one should also be sensitive to the presence of a Higgs boson that has up to 5\% of the
standard model cross section, has a mass as large as $140\,\rm{GeV}$ and decays invisibly. Because of the invisible decay, one 
needs an extra photon in the production. Together with the reduced production cross-section this means, that one
needs about a 1000 times larger luminosity than is normally considered in the discussion.

Therefore we must consider electron-positron colliders.
One studies the Higgs boson in the recoil process
$$ e^+ e^- \rightarrow Z~~H$$
independent of its decay, which could be invisible.
What are the requirements? We need an energy up to 300 GeV, covering the 
range up to the $ZZ$-threshold for which measurable signals are dominated by rare decays.
 Beyond this threshold the LHC is able to make detailed studies.
One needs very high precision, since one wants to be sensitive to the standard model width $(4\,\rm{MeV})$.
This puts extremely high demands on both the machine and the detectors.
In order to reconstruct the Higgs line-shape from the recoil spectrum one needs to determine the 
incoming energy in the collision to $O(1\,\rm{MeV})$ and one needs to construct the momentum of the outgoing
leptons from the $Z$-decay to $O(1\,\rm{MeV})$ as well. At the same time one still needs a high luminosity.

These demands make a linear collider problematic, as beam-strahlung is likely to wash out the precision
on the energy of the incoming electrons/positrons. For the detector one needs momentum measurements \newline
$\Delta p/p = 10^{-5}$. This is about a factor 20 better than what is presently discussed. Maybe nanotechnology
can help here. 

We are therefore left with a large circular collider as a preferred option.
This option is nowadays a subject of intense study under the name FCC (future circular collider) [22-24].
Having a large ring, one should consider putting in hadrons later as well.
The question of course is how large should the ring be. This should be determined by physics and not by politics.
Of course one can hope for new physics at a large energy scale, but there is no particular reason
for it to exist. One should therefore base the design on existing features of standard model physics
that are known to exist and should be tested. For the lepton collider this means measuring the Higgs width.
Naive quadratic scaling from LEP leads to a circumference of about 230 km, which actually could
be built in Fermilab, where it was studied under the name of VLEP [25]. However more detailed accelerator studies
are needed.
For the hadron collider option one should
be sensitive to sphaleron processes [26,27]. This would imply an energy in the $100\,\rm{TeV}$ range; a 230 km ring
should be sufficient for this. For such a machine to be built a supra-regional collaboration is needed.
One could imagine a structure like CERN, but with regions instead of countries as units. If China and the U.S.A.
would get involved in a more major way in high energy physics, at least the lepton collider could be built in a relatively short
time. One could go in steps in energy, first repeating  LEP at higher precision. However present day politics gives
no reason for optimism.

\section{Conclusion}
So now we can conclude. Yes, with the discovery of the Higgs boson
scientific history has been made, the results will be lasting.
This progress was only made possible through the painstaking hard and precise work
by experimental and theoretical physicists, whose work receives little attention
in the press or through prizes.
We can therefore be grateful to the EU for supporting such research through the
``HiggsTools" network and its predecessors ``HepTools" and ``Physics at Colliders".  
Fanciful and science-fiction like scenarios, that have been
abundantly with us and in the headlines, ultimately have played little or no role.
With a large circular collider a clear path for the future was sketched, where precision
physics will again be crucial.\\
\begin{center}
 \Large{Precision = Discovery !!}
\end{center}

\section*{Acknowledgements}
I thank Prof. E.~W.~N.~Glover for discussions and Prof. M.~Schumacher for a careful reading of the manuscript.

\end{document}